\documentclass[a4paper,12pt]{article}

\newcommand{\htr}{{\bf \beta}}
\newcommand{\ohtr}{\dot\htr}
\newcommand{\ndl}{{\bf B}}

\title{Search for an equation of motion of a classical pointlike charge}
\author{Marijan Ribari\v c and Luka \v Su\v ster\v si\v c\thanks{Corresponding author. Phone +386 1 477 3258; fax +386 1 423 1569; electronic address: \tt luka.sustersic@ijs.si\rm} \\Jo\v zef Stefan Institute, p.p.3000, 1001 Ljubljana, Slovenia}
\date{}

\begin{document}

\maketitle


\vskip 0.5 in
In 1892 H.A.~Lorentz started the search for a classical equation of motion for pointlike charged particles that takes into account the radiation reaction force. This search culminated in the Lorentz-Abraham-Dirac equation of motion, which is not satisfactory since it exhibits self-acceleration causing runaway solutions. In spite of ongoing efforts for more than a century, there is yet no acceptable classical equation of motion for a pointlike charge, cf.~the recent paper by Rohrlich \cite{Rohrlich1} and the comments about his proposal \cite{mi001, Rohrlich2, Zhidkov}. So, it is still an open question how to augment continuous classical electrodynamics with the physical concept of pointlike charged particles. The pointlike charge is presently only a common and handy computational device, which we generalized by the expansions in terms of co-moving moments of time-dependent, moving charges and currents \cite{mi003}.

This open question suggests that our present understanding of classical electrodynamics is actually not as complete as is usually taken for granted. There are actually many serious conceptual flaws and crack in our understanding of the consequences of electromagnetic forces between electric charges and currents.

To faciliate the search for an equation of motion for a pointlike charge, in our monograph \cite{miknjiga} we:
\begin{itemize}
\item{}clarified the issues by concisely formulating this open question,
\item{}gave a systematic collection of conditions which prospective equations have to meet, and of physical properties they are desired to possess, and
\item{}gathered and collated various theoretical concepts and results which might be \it useful and relevant \rm to anyone interested in discussing and proposing answers to this open question of theoretical physics.
\end{itemize}
To this end we considered about 80 references from 1903 to 1989 where some of the ideas, concepts, conclusions or formulae analogous or relevant to those considered in our monograph may be found. 

To obtain an equation of motion, generalizing Schott \cite{Schott} we introduced the acceleration four-momentum $\ndl$ implicitely defined by the relativistic differential energy-momentum balance equation
\begin{equation}
   \ohtr - (\ohtr\cdot\ohtr)\htr + \tau_0\gamma d\ndl /dt = {\bf f} \,,
   \label{spenacba}
\end{equation}
where: (i)~$\htr = (\gamma,\gamma{\bf v}/c)$, with $\gamma = (1 - |{\bf v}/c|)^{-1/2}$, is the four-velocity of the pointlike charge; (ii)~$\ohtr = \tau_0\gamma d\htr /dt$, is the acceleration four-vector, with $\tau_0 = q^2/6\pi\epsilon_0 mc^3$, $q$ and $m$ being the pointlike charge charge and mass, respectively; (iii)~${\bf f} = (\tau_0/mc)\gamma({\bf v}\cdot {\bf F}/c, {\bf F})$ is the external four-force, with ${\bf F}$ being the external force acting on the pointlike charge. 

If the acceleration four-momentum $\ndl$ is explicitly known and depends solely on the external four-force ${\bf f}$ and four-momentum $\htr$ but not on its derivatives, relation (\ref{spenacba}) could be used as a Newtonian equation of motion; and for $\ndl = -\ohtr$, it is just the Lorentz-Abraham-Dirac equation. We pointed out sixteen qualitative properties that a general equation of motion for pointlike charged particles ought to possess \cite{miknjiga, mi001}; in particular, \it it has to have more than two free parameters, \rm with unforeseeable mathematical implications.

Furthermore, we proposed a conceptually new mathematical model---differential relations describing the asymptotic behaviour of trajectories of classical pointlike charged particles in response to a small and slowly changing external force $f_{ext} = \omega F(\omega t)$, $\omega > 0$ \cite{mi002, miknjiga}. They take account of radiation, and of Dirac's and Bhabha's conditions about conservation of energy and linear momentum, and of angular and boost momenta. We can use them for describing, investigating and evaluating the dynamic behaviour of classical pointlike charged particles in response to a small and slowly changing external force to any desired order of $\omega$. But we can not use these differential relations as equations of motion for computing their trajectories; though we may transform each of them into an appropriate Newtonian equation of motion accurate up to the same order of $\omega$. We proposed that the Lorentz-Abraham-Dirac equation is not an equation of motion for computing trajectories, but just a differential relation of order $\omega^2$, the lowest order that still takes account of radiation.


\begin{thebibliography}{99}

\bibitem{Rohrlich1}F. Rohrlich, Phys. Lett. A \bf283 \rm(2001) 276.

\bibitem{mi001}M. Ribari\v c and L. \v Su\v ster\v si\v c, Phys. Lett A \bf295 \rm(2002) 318.

\bibitem{Rohrlich2}F. Rohrlich, Phys. Lett. A \bf295 \rm(2002) 320.

\bibitem{Zhidkov}A. Zhidkov, J. Koga, A. Sasaki and M. Uesaka, Phys. Rev. Lett. \bf88 \rm(2002) 185002; \hfill\break G. Compagno and F. Persico, J. Phys. A-Math. Gen. \bf35 \rm(2002) 3629; \hfill\break W.E. Baylis and J. Huschilt, Phys. Lett. A \bf301 \rm(2002) 7; \hfill\break F. Rohrlich, Phys. Lett. A \bf303 \rm(2002) 307; \hfill\break R. Rivera and D. Villarroel, Phys. Rev. E \bf66 \rm(2002) 046618; \hfill\break D. Villarroel, Phys. Rev. E \bf66 \rm(2002) 046624; \hfill\break R.F. O'Connell, Phys. Lett. A \bf313 \rm(2003) 491; \hfill\break T. Petrosky, G. Ordonez and I. Progogine, Phys. Rev. A \bf68 \rm(2003) 022107; \hfill\break D. Vogt and P.S. Letelier, Gen. Relativ. Gravit. \bf35 \rm(2003) 2261; \hfill\break J.A.E. Roa-Neri and J.L. Jimenez, Found. Phys. \bf34 \rm(2004) 547;
\hfill\break V.I. Berezhiani, R.D. Hazeltine and S.M. Mahajan, Phys. Rev. E \bf69 \rm(2004) 056406; \hfill\break R.D. Hazeltine and S.M. Mahajan, Phys. Rev. E \bf70 \rm(2004) 036404; \hfill\break R.D. Hazeltine and S.M. Mahajan, Phys. Rev. E \bf70 \rm(2004) 046407; \hfill\break J. Koga, Phys. Rev. E \bf70 \rm(2004) 046502; \hfill\break V.V. Lidsky, Theor. Math. Phys. \bf 143 \rm(2005) 583.

\bibitem{mi003}M. Ribari\v c and L. \v Su\v ster\v si\v c, SIAM J. Appl. Math. \bf 55 \rm (1995) 593.

\bibitem{miknjiga}M. Ribari\v c and L. \v Su\v ster\v si\v c, \it Conservation Laws and Open Questions of Classical Electrodynamics, \rm World Scientific, Singapore 1990, Chapters 9--11.

\bibitem{Schott}G.A. Schott, Ann. Phys. (Leipzig) \bf25 \rm (1908) 63.

\bibitem{mi002}M. Ribari\v c and L. \v Su\v ster\v si\v c, Phys. Lett. A\bf139 \rm(1989) 5.

\end{thebibliography}
\end{document}